\documentclass[prd,aps,tightenlines,superscriptaddress,amsmath,amssymb,amsfonts]{revtex4}
\usepackage{amssymb,amsmath,amsthm,amsbsy,epsfig,color,graphicx,times}
\usepackage[ansinew]{inputenc}
\usepackage[english]{babel}
\usepackage{color}
\usepackage{braket}
\usepackage{tabularx}
\usepackage{array}

\begin{document}

\title{Neutrinos in curved space-time: particle mixing and flavor oscillations }

\author{A. Capolupo}
\email{capolupo@sa.infn.it}
\affiliation{Dipartimento di Fisica ``E.R. Caianiello'' Universit\`{a} di Salerno, and INFN -- Gruppo Collegato di Salerno, Via Giovanni Paolo II, 132, 84084 Fisciano (SA), Italy}

\author{G. Lambiase}
\email{lambiase@sa.infn.it}
\affiliation{Dipartimento di Fisica ``E.R. Caianiello'' Universit\`{a} di Salerno, and INFN -- Gruppo Collegato di Salerno, Via Giovanni Paolo II, 132, 84084 Fisciano (SA), Italy}

\author{A. Quaranta}
\email{anquaranta@unisa.it}
\affiliation{Dipartimento di Fisica ``E.R. Caianiello'' Universit\`{a} di Salerno, and INFN -- Gruppo Collegato di Salerno, Via Giovanni Paolo II, 132, 84084 Fisciano (SA), Italy}

\begin{abstract}

We   present  a quantum field theoretical approach to the vacuum neutrino oscillations in curved space,
we   analyze  the non--trivial interplay between quantum field mixing and field quantization in curved space
 and derive new oscillation formulae.
We compute the formulae explicitly in the spatially flat FLRW metrics for universes dominated by a cosmological constant and by radiation. We evaluate the transition probabilities in the Schwarzschild black hole metric, and we show that the Hawking radiation affects the oscillations of neutrinos.
We show that our results are consistent with those of previous analyses when the quantum mechanical limit is considered.

\end{abstract}

\maketitle

\section{Introduction}

Since they were theoretically proposed by Pauli \cite{Pauli}, neutrinos have proven to be among the most enigmatic particles in the universe. Until the discovery of flavor oscillations~\cite{NeutrinoOscillations1,NeutrinoOscillations2}, whose theory was pioneered by Pontecorvo~\cite{Pontecorvo,Bilenky}, neutrinos were believed to be massless. Today it is accepted that neutrinos are massive particles, and that they oscillate among three flavors $\nu_e, \nu_{\mu}, \nu_{\tau}$ corresponding to the companion charged leptons $e, \mu, \tau$. This peculiarity renders neutrinos unique among the known elementary particles and puts them beyond the scope of the standard model of particles~\cite{Mohapatra}. In many respects, neutrinos are forerunners of a new physics, as several issues, including the origin of their mass~\cite{NeutrinoMass} and their fundamental nature~\cite{NeutrinoNature}, are still open to the present day.

On the other hand, the relevance of neutrinos in astrophyisical and cosmological contexts has grown dramatically during the last years. They figure as a valuable source of information, along with gravitational waves and electromagnetic radiation, in the ever--growing field of multi--messenger astronomy~\cite{Franckowiak}. The study of neutrinos of astrophysical origin can indeed provide fundamental insights on the source that produced them. In addition, neutrinos are expected to play an important role in the first phases of the universe~\cite{Buchmuller,CosmologicalNeutrinos}, and the detection of the cosmic neutrino background, pursued in experiments as PTOLEMY~\cite{PTOLEMY}, could represent an essential test for the standard cosmological model \cite{CosmologicalNeutrinos}. Mass varying neutrinos have also been proposed as a possible explanation for Dark Energy \cite{Mavans}.

This state of affairs requires a careful investigation of neutrino oscillations on a curved spacetime. The topic has been discussed in several works , where it was found that gravitational fields may alter both the oscillations in vacuum and in matter~\cite{Grossman,Piriz,Cardall}.

Here we wish to go beyond the heuristic treatment of ref.~\cite{Cardall}, and present a quantum field theoretical approach, based on the field quantization in curved space-time, to evaluate the effects of gravitational fields on neutrino oscillations.
We derive  general  oscillation formulae for flavor fields in curved space-time, which represent our main result.
We discuss the particle interpretation of the fields in presence of gravity and study how the mixing changes when moving from a mass field representation to another. We demonstrate the invariance of local observables, which are represented by expectation values on flavor states of local operators constructed from the flavor fields. We show that the oscillation probabilities, on the other hand, do in general
 depend on the representation of the mass fields, since they are not a local observable and involve the comparison between particles in different spacetime regions. We establish the conditions  which have to be satisfied in order that the resulting transition probabilities are invariant under changes of mass field representation.

We also compute explicitly the oscillation formulae for two examples of spatially flat Friedmann--Lemaitre--Robertson--Walker spacetimes, corresponding to a cosmological constant--dominated and a radiation--dominated universe respectively. In these cases, exact analytical solutions to the Dirac equation are available, and the formalism here introduced can be applied  directly.
Moreover, we give an estimation of the oscillation formulae for neutrinos propagating from the past infinity to the future infinity in a stationary Schwarzschild spacetime.
We  introduce a method to extract the oscillation formulae on spacetimes with asymptotically flat regions without resorting to the exact solutions of the Dirac equation. We then employ this strategy to compute the formulae on the Schwarzschild black hole spacetime, for neutrinos propagating from the past infinity to the future infinity. We show how the Hawking radiation is naturally embedded in the resulting transition probabilities.

Our results generalize those of the previous treatments ~\cite{Cardall}, and are consistent with the latter when the suitable limits are considered.
In our computation, for simplicity, we limit our analysis to the vacuum oscillations, therefore considering the sole effect of gravity.

The paper is organized as follows: in section II we provide the setting for the description of the mass fields in curved space; in section III we develop field mixing and find the oscillation probabilities in curved spacetime, with a thorough analysis of their features; in section IV we apply the formalism to some spacetimes of interest, including the spatially flat FLRW metric for a radiation-dominated universe and for a cosmological constant-dominated universe, and the Schwarzschild black hole metric, where we show the impact of the Hawking effect on neutrino oscillations; finally in section V we draw our conclusions.

\section{Mass neutrino fields in curved Space}

To evaluate the oscillation formulae for neutrinos on a curved spacetime, it is necessary to consider both the effects of curvature and mixing on the (free) mass fields. Let $M$ be a globally hyperbolic spacetime, and let $\tau \in \mathbb{R}$ label a foliation of $M$ by Cauchy surfaces. Consider the tetrad fields $e^{\mu}_{a} (x)$ satisfying $\eta^{a b} e^{\mu}_{a} (x) e^{\nu}_{b} (x) = g^{\mu \nu} (x)$. Here $\eta^{a b} \equiv diag(1,-1,-1,-1)$ is the Minkowski metric tensor, while $g^{\mu \nu}(x)$ is the contravariant metric $g^{\mu \nu} (x) g_{\nu \rho} (x) = \delta^{\mu}_{\rho}$ on $M$ in a given coordinate system. The massive neutrino fields satisfy the Dirac equations:
\begin{equation}\label{DiracEquation}
  (i \gamma^{\mu} (x) D_{\mu} - m_i) \psi_i = 0
\end{equation}
where $\gamma^{\mu} (x) = e^{\mu}_{a} (x) \gamma^{a}$, $\gamma^{a}$ being the usual flat space Dirac matrices, and $D_{\mu} = \partial_{\mu} - \frac{i}{4}\omega_{\mu}^{a b} \sigma_{a b}$. The spin connection is defined as $\omega_{\mu}^{a b} = e^{a}_{\nu} \Gamma^{\nu}_{\rho \mu}e^{\rho b}+e^{a}_{\nu} \partial_{\mu} e^{\nu b}$, whereas $\sigma_{a b}$ are the commutators of flat Dirac matrices $\sigma_{a b} = \frac{i}{2} [\gamma^{a}, \gamma^{b}]$. In equation \eqref{DiracEquation}, the index $i=1,2,...,N$ ranges over the number of neutrino species $N$. For the sake of simplicity we focus on the case $N=2$, though the generalization to $N=3$ is straightforward. In general equation \eqref{DiracEquation} cannot be solved exactly. Even if one is able to find exact solutions, these do not play the same prominent role as their flat spacetime counterpart. It is well--known, indeed, that the positive frequency solutions of equation  cannot be defined univocally, and that, consequently, there is no natural (nor unique) particle interpretation for the corresponding Quantum Field Theory ~\cite{Birrell,Wald}. Nevertheless, the canonical quantization of the Dirac field proceeds along the same lines as in  Minkowski spacetime.

To perform a field expansion, one must find a set of positive $\zeta_{k,i}$ and negative $\xi_{k,i}$ frequency solutions for each of the equations \eqref{DiracEquation}. In general the bipartition of the solutions to eq. \eqref{DiracEquation} makes sense only locally, while there is no natural global definition of positive and negative frequency modes. Anyway, one is free to choose a set of modes \cite{Note1} $\{\zeta_{k,i},\xi_{k,i}\}$ , deemed to be positive/negative frequency modes according to some specified observer, and expand the field with respect to them, provided that they form a complete (and orthonormal) set of solutions under the inner product
\begin{equation}\label{InnerProduct}
  (a_i,b_i) = \int_{\Sigma(\tau)} \sqrt{-g}d \Sigma_{\mu}(\tau) \bar{a}_i \gamma^{\mu}(x) b_i \
\end{equation}
with $a_i,b_i$ any solution to equation \eqref{DiracEquation} with mass $m_i$ and $\bar{b}_i = b^{\dagger}_i \gamma^0 (x)$. Here $d \Sigma^{\mu} (\tau) = n^{\mu} (\tau) dV_{\tau}$ denotes the volume element on the surface $\tau$ with unit timelike normal $n^{\mu} (\tau)$. This has to hold separately for each $i =1,2$. As it is easy to prove, for $a_i,b_i$ solutions of the (same) Dirac equation, the inner product \eqref{InnerProduct} does not depend on the hypersurface chosen for the integration. In particular, it is independent on the foliation by Cauchy hypersurfaces employed. The fields can then be expanded as
\begin{equation}\label{FreeFieldExpansion}
 \psi_i (x) = \sum_{k,s} \left( \gamma_{k,s;i} \zeta_{k,s;i} (x) + \epsilon_{k,s;i}^{\dagger} \xi_{k,s;i} (x) \right)
\end{equation}
with the operator coefficients $\gamma_{k,s,i}, \epsilon_{k,s,i}$ satisfying the usual canonical anticommutation relations, $k$ momentum index and $s$ helicity index. The annihilators are also required to anticommute for $i \neq j$, and, in particular $\{\gamma_{k,s,i}, \gamma^{\dagger}_{k,s,j} \} = \delta_{ij}$, $\{\epsilon_{k,s,i}, \epsilon^{\dagger}_{k,s,j} \} = \delta_{ij}$. In equation \eqref{FreeFieldExpansion} we prefer to keep any space-time dependence within the modes, for ease of treatment with a general metric. The expansions \eqref{FreeFieldExpansion} define the mass Hilbert space $\mathcal{H}_{m} = H_1 \otimes H_2$, which is constructed out of the vacuum $|0_{m} \rangle = |0_1 \rangle \otimes |0_2 \rangle$. Here $\ket{0_i}$ is defined, as usual, by  $\gamma_{k,s,i}\ket{0_i} = 0 = \epsilon_{k,s,i}\ket{0_i} $ for each $k,s,i$.

As hinted above, the field expansions \eqref{FreeFieldExpansion} are somewhat arbitrary, as opposed to the flat spacetime case, where there is no ambiguity in the definition of positive and negative frequency modes. Any other basis $\{\tilde{\zeta}_{k,s,i},\tilde{\xi}_{k,s,i}\}$ can be used to expand the fields $\psi_i = \sum_{k,s}( \tilde{\gamma}_{k,s,i} \tilde{\zeta}_{k,s,i} (x) + \tilde{\epsilon}_{k,s,i}^{\dagger} \tilde{\xi}_{k,s,i} (x))$. Since both the sets $\{\zeta_{k,s,i},\xi_{k,s,i}\}$ and $\{\tilde{\zeta}_{k,s,i},\tilde{\xi}_{k,s,i}\}$ form a basis for the space of solutions of eqs. \eqref{DiracEquation}, one can write the modes of a set in terms of the other, for each $i$:
\begin{eqnarray}\label{BogoliubovModes}
 \nonumber \tilde{\zeta}_{k',s',i} &=& \sum_{k,s} \left( \Gamma_{k',s';k,s;i}^{*} \zeta_{k,s,i} + \Sigma_{k' ,s';k, s; i}^{*} \xi_{k, s, i} \right) \\
  \tilde{\xi}_{k', s', i} &=& \sum_{k , s} \left( \Gamma_{k', s';k, s; i} \xi_{k,s,i} - \Sigma_{k', s';k, s;i} \zeta_{k,s,i} \right)
\end{eqnarray}
where $ \Gamma_{k', s'; k s; i} = ( \tilde{\zeta}_{k',s',i}, \zeta_{k,s,i}) = (\xi_{k,s,i}, \tilde{\xi}_{k',s',i})$ and $\Sigma_{k' ,s';k s;i} =( \tilde{\zeta}_{k',s',i},\xi_{k,s,i}) = -(\zeta_{k,s,i}, \tilde{\xi}_{k' s',i}) $. This is a fermionic Bogoliubov transformation, for which \mbox{$\sum_{q,r} \left( \Gamma_{k ,s; q ,r;i}^{*} \Gamma_{k',s' ;q, r;i} + \Sigma_{k, s; q, r;i}^{*} \Sigma_{k',s';q, r;i} \right) = \delta_{k,k'} \delta_{s,s'}$} for each $i$. The corresponding relation between the two sets of annihilators is given by
\begin{eqnarray}\label{Bogoliubov1}
 \nonumber \tilde{\gamma}_{k,s,i} &=& \sum_{k',s'} \left(\Gamma_{k, s; k' s';i} \gamma_{k', s',i} +  \Sigma_{k, s;k',s';i} \epsilon^{\dagger}_{k',s',i} \right)\\
  \tilde{\epsilon}_{k ,s,i} &=& \sum_{k', s'} \left( \Gamma_{k, s; k', s';i} \epsilon_{k' ,s',i} -  \Sigma_{k, s;k', s';i} \gamma^{\dagger}_{k' ,s',i} \right) \ .
\end{eqnarray}
It is often the case that the Bogoliubov coefficients $\Gamma_{k, s; k', s';i} , \Sigma_{k, s; k', s';i}$ can be written as $\Gamma_{k, s; k',s';i} = \delta_{k,k'} \delta_{s,s'} \Gamma_{k,i}$, $\Sigma_{k, s; k', s';i} = \delta_{k,k'} \delta_{s,s'} \Sigma_{k,i}$, with $\Gamma_{k,i}$ and $\Sigma_{k,i}$ depending on $k$ alone. In this occurrence, they admit the parametrization $\Gamma_{k,i} = e^{i \eta_{k,i}} \cos(\theta_{k,i})$ ,$\Sigma_{k,i} = e^{i \phi_{k,i}} \sin(\theta_{k,i})$, with $\eta_{k,i},\phi_{k,i}, \theta_{k,i}$ real functions of $k$. We remark that the Bogoliubov transformations \eqref{Bogoliubov1} can be recast in terms of the generators $J_i = e^{\sum_{k,k',s,s'} \left[ ( \lambda_{k,k',s,s',i}^* \gamma_{k, s,i}^{\dagger}\epsilon_{k' s',i}^{\dagger} - \lambda_{k,k',s,s',i} \epsilon_{k, s,i} \gamma_{k' ,s',i})\right] }$, with  $\lambda_{k,k',s,s',i} = Arctan(\frac{\Sigma_{k, s;k' ,s';i}}{\Gamma_{k ,s;k',s';i}})$, as
\begin{equation}\label{CurvatureGen}
  \tilde{\gamma}_{k,s,i} = J_i^{-1} \gamma_{k,s,i} J_i  \;,  \qquad  \qquad \ \  \tilde{\epsilon}_{k,s,i} = J_i^{-1} \epsilon_{k,s,i} J_i \ .
\end{equation}
The maps $J_i:\tilde{\mathcal{H}}_i \rightarrow \mathcal{H}_i $ interpolate between the Fock spaces $\mathcal{H}_i$ built from the $\gamma_{k,s,i},\epsilon_{k,s,i}$ and the Fock space $\tilde{\mathcal{H}}_i$ built from the $\tilde{\gamma}_{k,s,i},\tilde{\epsilon}_{k,s,i}$. In particular, one has for the vacuum states $|\tilde{0}_i\rangle = J_i^{-1} |0_i \rangle$. As for the untilded representation, the mass Hilbert space in the tilded representation is the tensor product $\tilde{\mathcal{H}}_m = \tilde{\mathcal{H}}_1 \otimes \tilde{\mathcal{H}}_2$. It is convenient to define a unique generator of Bogoliubov transformations $J: \tilde{\mathcal{H}}_m \longrightarrow \mathcal{H}_m$ on $\tilde{\mathcal{H}}_m$ as the tensor product $J = J_1 \otimes J_2$. Then
\begin{equation}\label{CurvatureGen2}
  \tilde{\gamma}_{k,s,i} = J^{-1} \gamma_{k,s,i} J \;,  \qquad  \qquad \ \ \tilde{\epsilon}_{k,s,i} = J^{-1} \epsilon_{k,s,i} J
\end{equation}
for $i = 1,2$.
The expansions of the two fields $\psi_1$ and $\psi_2$ must be compatible with each other, i.e., each of the modes $\zeta_{k,s,2},\xi_{k,s,2}$ must be obtainable from the corresponding modes $\zeta_{k,s,1},\xi_{k,s,1}$ by the substitution $m_1 \leftrightarrow m_2$, and vice--versa. In the context of mixing, this ensures that the same kind of particle, described by the same set of quantum numbers, is being mixed. This, of course, does not undermine the arbitrariness in the choice of the modes; these can be any complete set of solutions to the Dirac equation, provided that the same choice is made for the two fields.

\section{Neutrino mixing and oscillation formulae in curved space-time}

In this Section, we  show new   oscillation formulae for flavor fields in curved space-time and we present general considerations on the infinitely many unitarily inequivalent representations of the canonical anticommutation relations which characterize the quantization of mixed fields and of fields in curved space.

\subsection{Oscillation Formulae}

As discussed above, the QFT of free Dirac fields in curved space is characterized by infinitely many unitarily inequivalent representations of the canonical anticommutation relations. The phenomenon of mixing, even in Minkowski space, suffers from an analogous ambiguity, in that the flavor and the mass representations are unitarily inequivalent ~\cite{Capolupo1}. The effects of such inequivalence have been analyzed in flat space time \cite{Capolupo:2006et} and the possibility to reveal them in experimental setup has been recently proposed \cite{Capolupo:2019gmn}.  Let us start by fixing the mass field expansions \eqref{FreeFieldExpansion} and describe the mixing in a given representation of the mass fields. The flavor fields are defined as $\psi_{e} = \cos(\theta) \psi_1 +  \sin(\theta) \psi_2$ and $\psi_{\mu} = \cos(\theta)\psi_2 - \sin(\theta) \psi_1$ with $\theta$ the (2-flavor) mixing angle. Just like the Bogoliubov transformations \eqref{CurvatureGen2}, the rotation to flavor fields can be cast in terms of a generator $\mathcal{I}_{\theta} (\tau)$. This is given by
\begin{equation}\label{CurvedMixingGenerator}
  \mathcal{I}_{\theta} (\tau) =  e^{\theta [(\psi_1,\psi_2)_{\tau}-(\psi_2,\psi_1)_{\tau}]} \ .
\end{equation}
where the scalar products $(\psi_i,\psi_j)$ \emph{do} depend on the hypersurface chosen for the integration, since they are solutions to different Dirac equations.
Then, by definition, the flavor fields are expressed as
\begin{equation}
\psi_e = \mathcal{I}_{\theta}^{-1} (\tau) \psi_1 \mathcal{I}_{\theta} (\tau) \;,  \qquad  \qquad \ \  \psi_{\mu} = \mathcal{I}_{\theta}^{-1} (\tau) \psi_2 \mathcal{I}_{\theta} (\tau).
\end{equation}
If we let the generator \eqref{CurvedMixingGenerator} act on the mass annihilators, we obtain the flavor annihilators for  curved space
\begin{equation}
  \gamma_{k,s,e} (\tau) =  \mathcal{I}_{\theta}^{-1} (\tau) \gamma_{k,s,1} \mathcal{I}_{\theta} (\tau)=  \cos(\theta) \gamma_{k,s,1} + \sin(\theta)    \sum_{q,r} \bigg[\Lambda^*_{q,r;k,s}(\tau) \gamma_{q,r,2}   + \Xi_{q,r;k,s}(\tau) \epsilon_{q,r,2}^{\dagger}\bigg]\,.
\end{equation}
And similar for $\gamma_{k,s,\mu} (\tau), \epsilon_{k,s,e} (\tau), \epsilon_{k,s,\mu} (\tau)$. The Bogoliubov coefficients are provided by the inner products of the solutions to the \emph{curved space} Dirac equation with mass $m_1$ and $m_2$, that is, $\Lambda_{q,r;k,s} (\tau) = (\zeta_{q,r,2},\zeta_{k,s,1})_{\tau} = (\xi_{k,s,1},\xi_{q,r,2})_{\tau}$ and $\Xi_{q,r;k,s} (\tau) = (\zeta_{k,s,1},\xi_{q,r,2})_{\tau} = - (\zeta_{q,r,2},\xi_{k,s,1})_{\tau}$. The mixing coefficients always satisfy
{\small
\begin{equation}\label{FermionBog}
\sum_{q,r} \left(\Lambda_{k ,s; q ,r}^{*}(\tau) \Lambda_{k',s' ;q, r}(\tau) + \Xi_{k, s; q, r}^{*}(\tau) \Xi_{k',s';q, r} (\tau) \right)\! = \! \delta_{k,k'} \delta_{s,s'} \ .
\end{equation}}
Since the mass expansions are compatible, the mixing coefficients are often diagonal, namely of the form
\begin{eqnarray}\label{Compatibility}
 \nonumber \Lambda_{q,r;k,s} (\tau) &=& \delta_{q,k}\delta_{r,s} \Lambda_{k,s}(\tau) \\
 \Xi_{q,r;k,s} (\tau) &=& \delta_{q,k}\delta_{r,s} \Xi_{k,s}(\tau)
\end{eqnarray}
with $\Lambda_{k,s}(\tau)$, $\Xi_{k,s}(\tau)$ depending on $k$ and $s$ alone \cite{Note3}. Exceptions to this arise when we consider expansions of the mass fields in terms of modes labelled by the energy. In such a case, the mixing coefficients are non--diagonal and different from zero $\Lambda_{\omega,\omega'} \neq 0$ ,$\Xi_{\omega,\omega'} \neq 0$, once $\omega$ is fixed, only for a specific value of $\omega'$.
\begin{equation}
|\Lambda_{k, s} (\tau)|^2 + |\Xi_{k ,s} (\tau)|^2 = 1
\end{equation}
for each $k,s,\tau$, and
\begin{equation}
|\Lambda_{\omega; \omega'} (\tau)|^2 + |\Xi_{\omega ;\omega'} (\tau)|^2 = 1
\end{equation}
respectively.  The mass and the flavor representations are unitarily inequivalent. For each $\tau$ one has a distinct flavor Fock space $\mathcal{H}_f (\tau)$ defined by $\gamma_{e,\mu} (\tau) , \epsilon_{e,\mu} (\tau)$. The flavor vacuum $\ket{0_f (\tau)} = \mathcal{I}_{\theta}^{-1} (\tau) \ket{0_m}$ is a condensate of $\psi_1,\psi_2$ particle-antiparticle pairs.

In order to define the transition probabilities, we observe that the total Lagrangian is invariant under $U(1)$ gauge transformations. Therefore the total charge $Q = Q_1 + Q_2 = Q_e + Q_{\mu}$ is conserved \cite{Note4}, where $Q_i = \sum_{k,s} \left( \gamma_{k,s,i}^{\dagger} \gamma_{k,s,i} - \epsilon_{k,s,i}^{\dagger} \epsilon_{k,s,i} \right)$ for $i=1,2,e,\mu$. It is then meaningful to define the transition probabilities as
\begin{equation}\label{TransitionProb}
   P^{\rho \rightarrow \sigma}_{k,s} (\tau) = \sum_{q,r} \bigg(\langle \nu_{\rho,k,s} (\tau_0) | Q^{q,r}_{\sigma} (\tau) | \nu_{\rho,k,s} (\tau_0) \rangle
 - \langle 0_f(\tau_0) | Q^{q,r}_{\sigma} (\tau) |0_f (\tau_0) \rangle \bigg).
\end{equation}
Here $\rho,\sigma = e,\mu$, the state  $| \nu_{\rho,k,s} (\tau_0) \rangle = \gamma^{\dagger}_{k,s,\rho} (\tau_0) \ket{0_f(\tau_0)}$ is the state with a single neutrino of flavor $\rho$, momentum $k$ and helicity $s$ on the reference hypersurface $\tau = \tau_0$ . The second term on the rhs of \eqref{TransitionProb} is just the implementation of the normal ordering with respect to $\ket{0_f(\tau_0)}$. By construction $ P^{e \rightarrow e}_{k,s}(\tau) +   P^{e \rightarrow \mu}_{k,s}(\tau) = 1$ and $P^{\mu \rightarrow e}_{k,s}(\tau) +   P^{\mu \rightarrow \mu}_{k,s}(\tau) = 1$ for each $\tau$. A straightforward calculation yields, in the general case (accounting for both diagonal and non--diagonal mixing coefficients) the result
\begin{equation}\label{DefinitiveProbs01}
  P^{e \rightarrow \mu}_{k,s}(\tau)  =   2 \cos^{2}(\theta) \sin^{2}(\theta)
  \times  \bigg[1 - \sum_{q,r} \Re\bigg(\Lambda_{k,s;q,r}^{*} (\tau_0) \Lambda_{k,s;q,r} (\tau)
  +  \Xi_{k,s;q,r}^{*} (\tau_0) \Xi_{k,s;q,r} (\tau)  \bigg)\bigg] \ .
\end{equation}
Equation \eqref{DefinitiveProbs01} is the central result of the paper. When equations \eqref{Compatibility} hold, this reduces to
\begin{equation}\label{DefinitiveProbs}
   P^{e \rightarrow \mu}_{k,s}(\tau)  =   2 \cos^{2}(\theta) \sin^{2}(\theta)
  \times  \bigg[1 -  \Re\bigg(\Lambda_{k,s}^{*} (\tau_0) \Lambda_{k,s} (\tau)
  +  \Xi_{k,s}^{*} (\tau_0) \Xi_{k,s} (\tau)  \bigg)\bigg] \ .
\end{equation}
In both cases one has
\begin{eqnarray}
 P^{e \rightarrow e}_{k,s}(\tau) = 1 -  P^{e \rightarrow \mu}_{k,s}(\tau) .
\end{eqnarray}

\subsection{Mixing on a curved background and gravity-induced ambiguity in the particle interpretation}

Up to now, we have worked within a fixed, but arbitrary, representation of the mass fields. The question arises about the other possible representations, and how the mixing changes when moving from a representation to another. For the definition \eqref{TransitionProb} to make sense, we must determine if and how the probabilities vary when the mass representation is changed. We take as a guideline the principle of covariance, so that the \emph{local} physical observables should be independent of the underlying representation. In moving from a given representation $\{\gamma_1,\epsilon_1 \}, \{\gamma_{2}, \epsilon_2 \}$ to another $\{\tilde{\gamma}_1,\tilde{\epsilon}_1 \}, \{ \tilde{\gamma}_2, \tilde{\epsilon}_2 \}$, we know how to connect the mass Fock spaces, namely via the generator \eqref{CurvatureGen2} $J^{-1}:\mathcal{H}_m \rightarrow \tilde{\mathcal{H}}_{m}$. For each mass representation, we can proceed as we did above and build the corresponding flavor annihilators and flavor spaces $\mathcal{H}_{f}(\tau),\tilde{\mathcal{H}}_f (\tau)$, together with the mixing generators $\mathcal{I}_{\theta} (\tau) : \mathcal{H}_f (\tau) \rightarrow \mathcal{H}_m $, $\tilde{\mathcal{I}}_{\theta}(\tau): \tilde{\mathcal{H}}_f (\tau) \rightarrow \tilde{\mathcal{H}}_m$. It is useful, at this point, to determine the relations among the mixing coefficients $\Lambda(\tau), \Xi (\tau) $ and $\tilde{\Lambda}(\tau), \tilde{\Xi}(\tau) $ that appear in the explicit form of the two generators $\mathcal{I}_{\theta} (\tau)$ and $\tilde{\mathcal{I}}_{\theta} (\tau)$. By definition we have
\begin{equation}
  \tilde{\Lambda}_{q,r;k,s}(\tau) = (\tilde{\zeta}_{q,r;2}, \tilde{\zeta}_{k,s;1})_{\tau}
  = \sum_{q',k',r',s'}\bigg(\bigg[\Gamma_{q,r;q',r';2}^{*} \zeta_{q',r';2} + \Sigma_{q,r;q',r';2}^{*} \xi_{q',r';2}\bigg]  , \bigg[\Gamma_{k,s;k',s';1}^{*} \zeta_{k',s';1} + \Sigma_{k,s;k',s';1}^{*} \xi_{k',s';1}\bigg]\bigg)_{\tau}\,.
 \end{equation}
 Here the first equality is just the definition of   $\tilde{\Lambda} (\tau)$, the second follows from the Bogoliubov transformations \eqref{Bogoliubov1}. By using the properties of the inner product \eqref{InnerProduct}, in the general case (again, accounting for both the diagonal and non--diagonal mixing coefficients), we obtain
   \begin{eqnarray}
  \nonumber   \tilde{\Lambda}_{q,r;k,s}(\tau) &=& \sum_{q',k',r',s'} \bigg[ \Gamma_{q,r;q'r';2} \Gamma_{k,s;k',s';1}^{*} (\zeta_{q',r';2},\zeta_{k',s';1})_{\tau} + \ \ \Gamma_{q,r;q',r';2} \Sigma_{k,s;k',s';1}^{*} (\zeta_{q',r';2},\xi_{k',s';1})_{\tau} \\  &+& \ \ \Sigma_{q,r;q'r';2} \Gamma_{k,s;k's';1}^{*} (\xi_{q',r';2},\zeta_{k',s';1})_{\tau}  + \ \ \Sigma_{q,r;q',r';2} \Sigma_{k,s;k',s';1}^{*} (\xi_{q',r';2},\xi_{k',s';1})_{\tau} \bigg]\,,
  \end{eqnarray}
  and, finally, from the definition of $\Lambda(\tau)$ and $\Xi(\tau)$, we have
  \begin{eqnarray}  \label{MixingCoefficient1}
   \nonumber  \tilde{\Lambda}_{q,r;k,s}(\tau) =   \sum_{q',k',r',s'} \bigg[ \Gamma_{q,r,q'r';2}
     \left(\Gamma_{k,s;k',s';1}^{*} \Lambda_{q',r';k',s'}(\tau)-  \Sigma_{k,s;k',s';1}^{*} \Xi_{q',r';k',s'}(\tau) \right) \\
   + \ \Sigma_{q,r;q',r';2}
    \left( \Gamma_{k,s;k',s';1}^{*} \Xi_{q',r';k's'}^{*} (\tau)  +  \Sigma_{k,s;k',s';1}^{*} \Lambda_{q',r';k',s'}^{*}(\tau) \right) \bigg]
   \ .
\end{eqnarray}
 Similarly we have
\begin{eqnarray}\label{MixingCoefficient2}
 \nonumber \tilde{\Xi}_{q,r;k,s}(\tau) =  \sum_{q',k';r',s'} \bigg[ \Gamma_{q,r;q',r';2}
 \left(\Gamma_{k,s;k',s';1} \Xi_{q',r';k',s'}(\tau) +   \Sigma_{k,s;k',s';1} \Lambda_{q',r';k',s'}(\tau) \right) \\
   -  \ \Sigma_{q,r;q',r';2}
  \left( \Gamma_{k,s;k',s';1} \Lambda_{q',r';k',s'}^{*} (\tau) - \ \ \Sigma_{k,s;k',s';1} \Xi_{q',r';k',s'}^{*}(\tau) \right)\bigg] \
 .
 \end{eqnarray}
When equations \eqref{Compatibility} hold for both the representations $\{q,r\},\{q',r'\}$, the equations reduce to
  \begin{eqnarray}\label{MixingCoefficient3}
  \nonumber  \tilde{\Lambda}_{q,r}(\tau) &=&   \sum_{q',r'} \bigg[ \Gamma_{q,r,q'r';2}
     \left( \Gamma_{q,r;q',r';1}^{*} \Lambda_{q',r'}(\tau) -  \Sigma_{q,r;q',r';1}^{*} \Xi_{q',r'}(\tau) \right) \\
    &+& \ \Sigma_{q,r;q',r';2}
    \left(\Gamma_{q,r;q',r';1}^{*} \Xi_{q',r'}^{*} (\tau) +  \Sigma_{q,r;q',r';1}^{*} \Lambda_{q',r'}^{*}(\tau) \right)\bigg]
    .
\end{eqnarray}
and
\begin{eqnarray}\label{MixingCoefficient4}
 \nonumber  \tilde{\Xi}_{q,r}(\tau) &=&  \sum_{q',r'} \bigg[ \Gamma_{q,r;q',r';2}
  \left( \Gamma_{q,r;q',r';1} \Xi_{q',r'}(\tau) +   \Sigma_{q,r;q',r';1} \Lambda_{q',r'}(\tau) \right)\\
 &-& \   \Sigma_{q,r;q',r';2} \left( \Gamma_{q,r;q',r';1} \Lambda_{q',r'}^{*} (\tau)  -   \Sigma_{q,r;q',r';1} \Xi_{q',r'}^{*}(\tau) \right) \bigg]  \ .
\end{eqnarray}

The equations (\ref{MixingCoefficient1}, \ref{MixingCoefficient2}, \ref{MixingCoefficient3}, \ref{MixingCoefficient4}) provide an explicit relation between $\mathcal{I}_{\theta}(\tau)$ and $\tilde{\mathcal{I}}_{\theta}(\tau) $, and show how the mixing coefficients change, in moving from a mass representation to another, in order to ensure covariance. In particular, the tilded coefficients turn out to be a linear combination of the untilded coefficients weighted by the coefficients of the Bogoliubov transformations between the two mass representations. A slightly modified version of the eqs. \eqref{MixingCoefficient1} and \eqref{MixingCoefficient2} will be expedient in the calculation of the transition probabilities in a number of interesting cases.

It remains to establish how the flavor operators $\gamma_{\rho} (\tau), \epsilon_{\rho} (\tau)$ and the flavor vacuum $|0_f (\tau) \rangle$ transform under a change of mass representation. We focus on the vacuum state $|0_f (\tau) \rangle \in \mathcal{H}_f (\tau)$. First we employ the generator $\mathcal{I}_{\theta} (\tau) : \mathcal{H}_f (\tau) \rightarrow \mathcal{H}_{m} $ to get the mass vacuum $|0_m \rangle$. Then we apply the generator of Bogoliubov transformations \eqref{CurvatureGen2} $J^{-1}:\mathcal{H}_{m} \rightarrow \tilde{\mathcal{H}}_{m}$ to obtain $|\tilde{0}_{m} \rangle$. Finally, the generator of mixing in the tilde representation $\tilde{\mathcal{I}}_{\theta}^{-1} (\tau) : \tilde{\mathcal{H}}_m \rightarrow \tilde{\mathcal{H}}_f (\tau)$ is employed to get $|\tilde{0}_f (\tau) \rangle $. We conclude that the two flavor vacua are related by the transformation
\begin{equation}\label{FlavorTransform}
 |\tilde{0}_f (\tau) \rangle  = J_f^{-1} (\tau) |0_f (\tau) \rangle \doteq \tilde{\mathcal{I}}_{\theta}^{-1} (\tau) J^{-1} \mathcal{I}_{\theta} (\tau) |0_f (\tau) \rangle \
\end{equation}
where we have defined the inverse $J_f^{-1} (\tau)$ for convenience. The flavor operators must then transform as
\begin{eqnarray}\label{OpFlavorTransform}
\nonumber \gamma_{k,s,\rho}(\tau) &\rightarrow& J_f^{-1}(\tau)  \gamma_{k,s,\rho}(\tau) J_f (\tau)  \\ \epsilon_{k,s,\rho}(\tau) &\rightarrow& J_f^{-1}(\tau)  \epsilon_{k,s,\rho}(\tau) J_f (\tau)
\end{eqnarray}
and similarly for the creation operators. Equations \eqref{FlavorTransform} and \eqref{OpFlavorTransform} ensure the invariance of local observables in the form of expectation values $\bra{\psi_f (\tau)} F(\psi_e(\tau), \psi_{\mu}(\tau)) \ket{\psi_f(\tau)}$ with $\ket{\psi_{f}(\tau)} \in \mathcal{H}_f (\tau)$ and $F(\psi_e(\tau), \psi_{\mu}(\tau))$ any operator constructed from the fields $\psi_e(\tau), \psi_{\mu}(\tau)$.

\subsection{Transition probabilities and the mass representation}

The oscillation probabilities are not a local observable, since they involve the comparison between particles at different values of $\tau$, as it is evident from the definition \eqref{TransitionProb}. In general these quantities \emph{do} depend on the representation of the mass fields, and this is because distinct representations might assign a different meaning to the quantum numbers $k,s$. For example, one might consider two expansions of the mass fields, one in terms of plane waves, labelled by the three--momenta $\{\pmb{k}\}$, and one in terms of localized wave packets, labelled by a suitable set of quantum numbers $\{q\}$. It is clear that the two expansions describe particles with different physical properties; the first describes particles with definite momentum, the second describes particles for which momentum and position are definite to some extent. Therefore the probabilities $P^{\rho \rightarrow \sigma}_{k,s}$ and $P^{\rho \rightarrow \sigma}_{q,s}$ refer to the oscillations of different particles, and have a different interpretation. It would make no sense, in such a case, to require the equivalence of the two. This, of course, would be true even in flat space.

What is meaningful to require, is that the transition probabilities $P^{\rho \rightarrow \sigma}_{k,s}$ be the same for each compatible representation, i.e., for each representation that refers to the same kind of particle, and therefore agrees on the meaning of the quantum numbers $k,s$. In mathematical terms, any two such representations shall be connected by diagonal Bogoliubov transformations
\begin{eqnarray}\label{BogoliubovModes2}
 \nonumber \tilde{\zeta}_{k,s,i} &=&  \Gamma_{k,s;i}^{*} \zeta_{k,s,i} + \Sigma_{k, s; i}^{*} \xi_{k, s, i} \\
  \tilde{\xi}_{k, s, i} &=&  \Gamma_{k, s; i} \xi_{k,s,i} - \Sigma_{k, s;i} \zeta_{k,s,i} \
\end{eqnarray}
 where it is understood that $\Gamma_{k,s;q,r;i} = \delta_{k,q} \delta_{s,r} \Gamma_{k,s;i}$ and $\Sigma_{k,s;q,r;i} = \delta_{k,q} \delta_{s,r} \Sigma_{k,s;i}$,
 In this case the transition probabilities $P^{\rho \rightarrow \sigma}_{k,s}$ are indeed the same, and this can be proven explicitly, by writing out equation \eqref{DefinitiveProbs} for the two representations
\begin{eqnarray}\label{DefinitiveProbs1}
   P^{e \rightarrow \mu}_{k,s}(\tau)  =   2 \cos^{2}(\theta) \sin^{2}(\theta)
  \bigg[1 - \sum_{q,r} \Re\bigg(\Lambda_{k,s;q,r}^{*} (\tau_0) \Lambda_{k,s;q,r} (\tau)
    + &\Xi_{k,s;q,r}^{*} (\tau_0) \Xi_{k,s;q,r} (\tau)  \bigg)\bigg]
\end{eqnarray}
and
\begin{eqnarray}\label{DefinitiveProbs2}
  \tilde{P}^{e \rightarrow \mu}_{k,s}(\tau)  =   2 \cos^{2}(\theta) \sin^{2}(\theta)
  \times  \bigg[1 - \sum_{q,r} \Re\bigg(\tilde{\Lambda}_{k,s;q,r}^{*} (\tau_0) \tilde{\Lambda}_{k,s;q,r} (\tau)
   +  \tilde{\Xi}_{k,s;q,r}^{*} (\tau_0) \tilde{\Xi}_{k,s;q,r} (\tau)  \bigg)\bigg]
\end{eqnarray}
With the aid of equations \eqref{MixingCoefficient1} , \eqref{MixingCoefficient2} and \eqref{BogoliubovModes2} we find

\begin{eqnarray}
\nonumber && \tilde{\Lambda}_{k,s;q,r}^{*} (\tau_0) \tilde{\Lambda}_{k,s;q,r} (\tau) +  \tilde{\Xi}_{k,s;q,r}^{*} (\tau_0) \tilde{\Xi}_{k,s;q,r} (\tau) = \\
 \nonumber && + \left (\Lambda_{k,s;q,r}^{*} (\tau_0) \Lambda_{k,s;q,r} (\tau) + \Xi_{k,s;q,r}^{*} (\tau_0) \Xi_{k,s;q,r} (\tau) \right)
  \left[ |\Gamma_{k,s,2}|^2 |\Gamma_{q,r,1}|^2  + |\Gamma_{k,s,2}|^2 |\Sigma_{q,r,1}|^2 \right]  \\
  && + \left(\Lambda_{k,s;q,r} (\tau_0) \Lambda_{k,s;q,r}^{*} (\tau) + \Xi_{k,s;q,r} (\tau_0) \Xi_{k,s;q,r}^* (\tau) \right)
 \left[ |\Sigma_{k,s,2}|^2 |\Gamma_{q,r,1}|^2  + |\Sigma_{k,s,2}|^2 |\Sigma_{q,r,1}|^2 \right] \ .
\end{eqnarray}
Each of the terms in the square brackets is real. Considered that
\begin{eqnarray}
 \Lambda_{k,s;q,r} (\tau_0) \Lambda_{k,s;q,r}^{*}(\tau)  =  \left(\Lambda_{k,s;q,r}^{*} (\tau_0) \Lambda_{k,s;q,r} (\tau) \right)^*\,, \qquad \Xi_{k,s;q,r} (\tau_0) \Xi_{k,s;q,r}^{*}(\tau) = \left(\Xi_{k,s;q,r}^{*} (\tau_0) \Xi_{k,s;q,r} (\tau) \right)^*\,,
\end{eqnarray}
and that the Bogoliubov coefficients satisfy $|\Gamma_{k,s,i}|^2 + |\Sigma_{k,s,i}|^2 =1$ for each $k,s,i$, we finally get

\begin{eqnarray}\label{PartialInvariance}
\Re[\tilde{\Lambda}_{k,s;q,r}^{*} (\tau_0) \tilde{\Lambda}_{k,s;q,r} (\tau) +  \tilde{\Xi}_{k,s;q,r}^{*} (\tau_0) \tilde{\Xi}_{k,s;q,r} (\tau)] =
   \Re[\Lambda_{k,s;q,r}^{*} (\tau_0) \Lambda_{k,s;q,r} (\tau) +  \Xi_{k,s;q,r}^{*} (\tau_0) \Xi_{k,s;q,r} (\tau)] \,,
\end{eqnarray}
which proves the invariance of \eqref{DefinitiveProbs1}.

In the most general case, as the quantum numbers $k,s$ and $k',s'$ have a different phyiscal meaning, the probabilities $P^{\rho \rightarrow \sigma}_{k,s}$ and $\tilde{P}^{\rho \rightarrow \sigma}_{k',s'}$ have different interpretations. Different representations of the mass fields do indeed assign a different meaning to such indices, so that the probabilities \eqref{TransitionProb} have no invariant meaning.
In order to make sense of the probabilities in eqs. \eqref{TransitionProb} in the most general case, a representation of the mass fields must be fixed on the grounds of physical relevance.
When the underlying spacetime $M$ possesses non trivial symmetries, as time translational invariance or spherical symmetry, there is no doubt that the representation should be fixed so to take them into account. In these cases "good quantum numbers" are suggested by the symmetries themselves (for instance, the energy $\omega$ for stationary metrics, the angular momentum $l,m$ for spherically symmetric spacetimes). In any case, the probabilities in a given mass representation can always be related to the probabilites in any other mass representation with the aid of equations \eqref{MixingCoefficient3} and \eqref{MixingCoefficient4}.
As a final remark, we stress that the issue discussed here has nothing to do with the diffeomorphism covariance of the theory. All the probabilities \eqref{DefinitiveProbs} are (generally covariant) scalars, as it is evident from the definitions.

\section{Neutrino oscillation formulae in FLRW metrics and in presence of a Schwarzschild black hole}

In this section we apply the formalism developed above to some cases of interest. After an analysis of the flat space limit, we consider two cosmologically relevant FLRW metrics, corresponding to a cosmological constant--dominated and a radiation--dominated universe respectively. In these cases, exact analytical solutions to the Dirac equation are available, and it is possible to employ equation \eqref{DefinitiveProbs01} directly. We then introduce a method to extract the oscillation formulae on spacetimes with asymptotically flat regions without resorting to the exact solutions of the Dirac equation. We employ this strategy to compute the formulae on the Schwarzschild black hole spacetime, for neutrinos propagating from the past infinity to the future infinity. We show how the Hawking radiation is naturally embedded in the resulting transition probabilities.

\subsection{Flat spacetime limit}

   As a first, trivial, application of the formulae \eqref{DefinitiveProbs}, let us check the flat spacetime limit. We can see at once that the equations \eqref{DefinitiveProbs} reduce to the ordinary oscillation formulae. Indeed, in this case, we can choose the cauchy hypersurfaces to be the $t = constant$ surfaces in a given Minkowskian coordinate system, while the modes $\{\zeta_{\pmb{k},s,i} (x), \xi_{\pmb{k},s,i}(x) \}$ are just the plane wave solutions to the flat Dirac equation with definite momentum, so that $\Lambda_{q,r;k,s} (t) \rightarrow U_{q,r;k,s} (t)$ and $\Xi_{q,r;k,s} (t) \rightarrow V_{q,r;k,s} (t)$, where $U_{q,r;k,s}$ and $V_{q,r;k,s}$ are the usual mixing coefficients in flat space \cite{Capolupo1}. Assuming, without loss of generality, $k$ along the $z$ direction, the helicity indices decouple $U_{q,r;k,s} = \delta^3 (\pmb{k}-\pmb{q}) \delta_{r,s} U_{\pmb{k}}$, $V_{q,r;k,s}=\delta^3 (\pmb{k}-\pmb{q}) \delta_{r,s} (-1)^s V_{\pmb{k}}$. Since $U_{\pmb{k}}(t) = U_{\pmb{k}}(0)  e^{i(\omega_{k,2} - \omega_{k,1})t}$ and $V_{\pmb{k}}(t) = V_{\pmb{k}}(0) e^{i(\omega_{k,2} + \omega_{k,1})t}$, we get

\begin{eqnarray}
\nonumber \sum_{k',r',q,r}\Re[U^{*}_{k,s;k',r'}(0) U_{k',r';q,r}(t) &+& V^{*}_{k,s;k',r'} (0) V_{k',r';q,r}(t)]
\; = \; \Re [|U_{\pmb{k}}(0)|^2 e^{i(\omega_{k,2} - \omega_{k,1})t} + |V_{\pmb{k}}(0)|^2e^{i(\omega_{k,2} + \omega_{k,1})t}  ]
\\
&=& |U_{\pmb{k}}(0)|^2 \cos [(\omega_{k,2} - \omega_{k,1})t] +  |V_{\pmb{k}}(0)|^2 \cos [(\omega_{k,2} + \omega_{k,1})t].
\end{eqnarray}

Substituting this result in eqs. \eqref{DefinitiveProbs} yields the flat space oscillation formulae, which further reduce to the Pontecorvo oscillation formulae in the quantum mechanical limit ($V_{\pmb{k}} = 0$).
Flat space also offers the possibility to illustrate some points discussed above in the simplest possible context. One might well expand the mass fields in terms of modes with definite energy and angular momentum $\zeta_{\omega,\kappa_j,m_j;i}$, $\xi_{\omega,\kappa_j,m_j,s;i}$ instead of considering modes with definite cartesian three--momentum $\pmb{k}$. The former shall be suitable combinations of spherical spinors \cite{AngularMom}. An interesting aspect, is that in such a representation, the mixing coefficients are no longer diagonal. Indeed one has
\begin{equation}\label{MixingCoeffSpher}
  \Lambda_{\omega' \kappa'_j m'_j ; \omega \kappa_j m_j}(t) =  \delta_{\omega', \sqrt{\omega^2 + \Delta m^2} } \delta_{\kappa'_j,\kappa_j} \delta_{m'_j,m_j} |U_{\omega, \omega'}| e^{i(\omega' - \omega)t}
\end{equation}
with $\Delta m^2 = m_2^2 - m_1^2$ and
{\small
\begin{equation}\label{MixingCoeffSpher1}
|U_{\omega, \omega'}| = \sqrt{\frac{\omega' + m_2}{2 \omega'}}\sqrt{\frac{\omega + m_1}{2 \omega}} \left(1 + \sqrt{\frac{(\omega' - m_2)(\omega - m_1)}{(\omega + m_1)(\omega' + m_2)}} \right) \ ,
\end{equation}}
and similar for $\Xi$, where the exponential is $e^{i(\omega' + \omega)t}$ and $|U_{\omega,\omega'}|$ is replaced by
{\small
\begin{equation}\label{MixingCoeffSpher2}
  |V_{\omega, \omega'}| = \sqrt{\frac{\omega' + m_2}{2 \omega'}}\sqrt{\frac{\omega + m_1}{2 \omega}} \left(\sqrt{\frac{\omega' - m_2}{\omega' + m_2}}- \sqrt{\frac{\omega - m_1}{\omega + m_1}} \right) \ .
\end{equation}}
Here the quantum number $\kappa_j$ refers to a relativistic generalization of the spin--orbit operator, which enters the Dirac equation in spherical coordinates \cite{AngularMom,Thaller}, and takes into account both the orbital and spin angular momentum. The index $m_j$ refers to one component of the total angular momentum $\pmb{J}$, and has not to be confused with the masses $m_i$. Without delving into the details of the calculation, the result of equation \eqref{MixingCoeffSpher} can be understood as follows. The modes, apart from a normalization constant, are given by
\begin{equation}\label{inmodes1}
\zeta_{\omega , \kappa_j ,m_j ; i}  = e^{- i \omega t } \sqrt{\frac{\omega + m_i}{2 \omega r^2}}
\begin{pmatrix}
 P_{\kappa_j}(\lambda_i r)\Omega_{\kappa_j,m_j} (\theta,\phi) \\
\sqrt{ \frac{\omega - m_i}{\omega + m_i}} P_{\kappa_j}(\lambda_i r)\Omega_{-\kappa_j,m_j} (\theta,\phi)
 \end{pmatrix}
\end{equation}

\begin{equation}\label{inmodes2}
\xi_{\omega , \kappa_j, m_j  ; i}  =e^{ i \omega t } \sqrt{\frac{\omega + m_i}{2 \omega r^2}} \begin{pmatrix}
 - \sqrt{ \frac{\omega - m_i}{\omega + m_i}} P_{\kappa_j}(-\lambda_i r)  \Omega_{\kappa_j,m_j} (\theta,\phi) \\
 P_{\kappa_j}(-\lambda_i r)\Omega_{-\kappa_j,m_j} (\theta,\phi)
 \end{pmatrix}
\end{equation}
where $\Omega_{\kappa_j,m_j} (\theta,\phi)$ are spherical spinors, and $ P_{\kappa_j}(\lambda_i r)$ are radial functions of the product $\lambda_i r$, with the radial momentum $\lambda_i = \sqrt{\omega^2 - m_i^2}$. These are solutions to the radial part of the Dirac equation, which turns out to be a Riccati--Bessel equation\cite{Abramowitz}. The functions $ P_{\kappa_j}(\lambda_i r)$ are combinations of spherical bessel functions $j_{n}$ of the form $P_{\kappa_j}(\lambda_i r) = r j_{\kappa_j} (\lambda_i r)$. In computing the inner products, the radial integration $\int dr P^{*}_{\kappa_j; 2}(\lambda_2 r)P_{\kappa_j; 1} (\lambda_1 r)  $ will produce a factor $\delta_{\lambda_2, \pm \lambda_1}$, because of the closure relation satisfied by the spherical Bessel functions. Since $\lambda_2 = \sqrt{\omega^{' 2} - m_2^2}$ and $\lambda_1 = \sqrt{\omega^2 - m_1^2}$, this will give rise to the delta factor appearing in  \eqref{MixingCoeffSpher}.
Notice that $|U_{\omega,\omega'}|, |V_{\omega,\omega'}|$ are numerically the same as the usual flat space coefficients $|U_{\pmb{k}}|,|V_{\pmb{k}}|$, when $k^2 = \omega^2 - m_1^2 = \omega^{'2} - m_2^2$ .This shows why the mixing coefficients $\Lambda$, $\Xi$ are not generally diagonal, and the flexibility of the formalism we have employed. Indeed, the non--diagonal coefficients automatically ensure that the flavor operators $\gamma_{\omega,\rho}$, $\epsilon_{\omega, \rho}$ take into account the mass difference, involving operators with distinct energies $\omega, \omega'$ for the fields $\psi_1$ and $\psi_2$ \cite{Note5}.
In flat space, the shift between the two representations is actually of no use. However, in a non--trivial framework, the versatility of the formalism is essential, as there are instances in which the cartesian components of the momentum $k_x,k_y,k_z$ are useless, while the "spherical" quantum numbers $\omega,l,m$ are well defined.

\subsection{Expanding universe with exponential growth of the scale factor}

 The simplest non--trivial application is to spatially flat Friedmann--Lemaitre--Robertson--Walker (FLRW) spacetimes. Consider the metric $ds^2 = dt^2 - a^2 (t) (dx^2 + dy^2 + dz^2)$ with an exponential expansion $a(t) = e^{Ht}$, $H = constant$. This is well--suited to describe a homogeneous, spatially flat and isotropic universe dominated by a cosmological constant. The normalized solutions to the Dirac equation for this metric were derived in \cite{Barut}. Assuming, without loss of generality, the momentum $k$ to be along the $z$ direction, the helicities decouple as in flat space. Choosing the Cauchy surfaces as the surfaces with $t=constant$, or equivalently with $a(t) = constant$, the mixing coefficients read
{ \small
\begin{eqnarray}\label{FRLWCOEFF1}
  &\Lambda_{k,s;q,r}& (t) = \delta_{s,r} \delta^3(\pmb{k}-\pmb{q})   \frac{\pi k e^{-Ht}}{2 H \sqrt{\cos(\frac{i \pi m_2}{H}) \cos(\frac{i \pi m_1}{H})}}
\times    \left[ \!  J_{v_1}^*   \! \! \left( \! \! \frac{k}{H}e^{-Ht} \! \! \right) \! \!  J_{v_2} \! \!  \left( \! \! \frac{k}{H}e^{-Ht} \! \! \right) \! +\! J_{v_1 - 1}^*\! \! \left(\! \!  \frac{k}{H}e^{-Ht}\! \!  \right) \! \!  J_{v_2-1} \! \!  \left(\! \!  \frac{k}{H}e^{-Ht}\! \!  \right)\! \right]  \\
&\Xi_{k,s;q,r}& (t) = \delta_{s,r} (-1)^s \delta^3(\pmb{k}-\pmb{q}) \frac{\pi k e^{-Ht}}{2 H \sqrt{\cos(\frac{i \pi m_2}{H}) \cos(\frac{i \pi m_1}{H})}}
 \left[ \!  J_{v_1}^*   \! \! \left( \! \! \frac{k}{H}e^{-Ht} \! \! \right) \! \!  J_{-v_2} \! \!  \left( \! \! \frac{k}{H}e^{-Ht} \! \! \right) \! +\! J_{v_1 - 1}^*\! \! \left(\! \!  \frac{k}{H}e^{-Ht}\! \!  \right) \! \!  J_{1-v_2} \! \!  \left(\! \!  \frac{k}{H}e^{-Ht}\! \!  \right)\! \right]
\end{eqnarray} }
where $J_{\alpha}$ denotes the $\alpha$ Bessel function and $v_j = \frac{1}{2} \left(1 + \frac{2im_j}{H} \right)$ for $j=1,2$. Plugging these expressions in equations \eqref{DefinitiveProbs}, we obtain
{\footnotesize
\begin{eqnarray}\label{ExponentialProbs}
  && P^{e \rightarrow \mu}_{k,s}(t) =  2 \cos^{2}(\theta) \sin^{2}(\theta)\bigg\{1 -  \frac{\pi^2 k^2e^{-H(t+t_0)}}{4 H^2 \cos(\frac{i \pi m_2}{H}) \cos(\frac{i \pi m_1}{H})}  \\ \nonumber
&& \times \Re \bigg[ \left[ \!  J_{v_1}   \! \! \left( \! \! \frac{k}{H} e^{-Ht_0} \! \! \right) \! \!  J_{v_2}^* \! \!  \left( \! \! \frac{k}{H}  e^{-Ht_0}\! \! \right) \! +\! J_{v_1 - 1}\! \! \left(\! \!  \frac{k}{H} e^{-Ht_0}\! \!  \right) \! \!  J_{v_2-1}^*  \! \!  \left(\! \!  \frac{k}{H} e^{-Ht_0}\! \!  \right)\! \right]
\left[ \!  J_{v_1}^*   \! \! \left( \! \! \frac{k}{H}e^{-Ht} \! \! \right) \! \!  J_{v_2} \! \!  \left( \! \! \frac{k}{H}e^{-Ht} \! \! \right) \! +\! J_{v_1 - 1}^*\! \! \left(\! \!  \frac{k}{H}e^{-Ht}\! \!  \right) \! \!  J_{v_2-1} \! \!  \left(\! \!  \frac{k}{H}e^{-Ht}\! \!  \right)\! \right]
\\ \nonumber && + \left[ \!  J_{v_1}   \! \! \left( \! \! \frac{k}{H} e^{-Ht_0} \! \! \right) \! \!  J_{-v_2}^* \! \!  \left( \! \! \frac{k}{H} e^{-Ht_0} \! \! \right) \! +\! J_{v_1 - 1}\! \! \left(\! \!  \frac{k}{H} e^{-Ht_0}\! \!  \right) \! \!  J_{1-v_2}^* \! \!  \left(\! \!  \frac{k}{H} e^{-Ht_0}\! \!  \right)\! \right]
  \left[ \!  J_{v_1}^*   \! \! \left( \! \! \frac{k}{H}e^{-Ht} \! \! \right) \! \!  J_{-v_2} \! \!  \left( \! \! \frac{k}{H}e^{-Ht} \! \! \right) \! +\! J_{v_1 - 1}^*\! \! \left(\! \!  \frac{k}{H}e^{-Ht}\! \!  \right) \! \!  J_{1-v_2} \! \!  \left(\! \!  \frac{k}{H}e^{-Ht}\! \!  \right)\! \right] \bigg] \bigg\}
\end{eqnarray}}

\subsection{Expanding universe dominated by radiation}

 Here we consider the FLRW metric for a radiation dominated universe $a(t) = a_0 t^{\frac{1}{2}}$. Notice that since $a(t)$ has to be adimensional, $a_0$ has dimension $[t]^{-\frac{1}{2}} = [m]^{\frac{1}{2}}$. As before, without loss of generality, we assume the neutrino momentum $k$ along the $z$ direction to decouple the helicities, and consider a foliation by the $t = constant$ hypersurfaces. The solutions to the Dirac equation for this metric are again found in \cite{Barut}, and yield the mixing coefficients
{\small
\begin{eqnarray}
  && \Lambda_{k,s;q,r} (t) = \delta_{s,r} \delta^3(\pmb{k}-\pmb{q}) \frac{1}{\sqrt[4]{4m_1m_2t^2}} e^{-\frac{\pi k^2 (m_1+m_2)}{4m_1m_2a_0^2}}
 \bigg\{ W^{*}_{\kappa_2,\frac{1}{4}} (-2im_2t) W_{\kappa_1,\frac{1}{4}} (-2im_1t)\\
 \nonumber && + \frac{4}{m_1 m_2 a_0^2 t}  \bigg[W^{*}_{\kappa_2, \frac{1}{4}} (- 2 im_2t) - \frac{1}{8}\left(1 - \frac{ik^2}{m_2 a_0^2} \right)W^{*}_{\kappa_2 - 1,\frac{1}{4}} (-2 i m_2 t) \bigg]  \bigg[ W_{\kappa_1,\frac{1}{4}}(-2 im_1t) - \frac{1}{8} \left(1 - \frac{ik^2}{m_1 a_0^2} \right)W_{\kappa_1 -1,\frac{1}{4}}(-2 i m_1 t) \bigg] \bigg\}
\end{eqnarray}}
{\small
\begin{eqnarray}
  && \Xi_{k,s;q,r} (t) = \delta_{s,r} \delta^3(\pmb{k}-\pmb{q}) \frac{k}{\sqrt[4]{2m_1(2m_2)^3a_0^2t}} e^{-\frac{\pi k^2 (m_1+m_2)}{4m_1m_2a_0^2}}  \bigg\{ W^{*}_{\kappa_1,\frac{1}{4}} (-2im_1t) W_{-\kappa_2,\frac{1}{4}} (-2im_2t)\\
\nonumber && + \frac{1}{m_1 m_2 a_0^2 t}
 \bigg[W^{*}_{\kappa_1, \frac{1}{4}} (- 2 im_1t) - \frac{1}{8}\left(1 - \frac{ik^2}{m_1 a_0^2} \right)W^{*}_{\kappa_1 - 1,\frac{1}{4}} (-2 i m_1 t) \bigg]   \bigg[ W_{-\kappa_2,\frac{1}{4}}(2 im_2t) + \frac{2im_2a_0^2}{k^2} W_{-\kappa_2 +1,\frac{1}{4}}(2 i m_2 t) \bigg] \bigg\}
\end{eqnarray}}
where $W_{\kappa, \mu} (z)$ are the Whittaker functions \cite{Abramowitz} and \mbox{$\kappa_j = \frac{1}{4} \left(1 + \frac{2ik^2}{a_0^2 m_j} \right)$ for $j=1,2$}. Insertion in eqs. \eqref{DefinitiveProbs} gives the transition probabilities ($t_0,t > 0$)
{\footnotesize
\begin{eqnarray}\label{WhittakerProbs}
\nonumber && P^{e \rightarrow \mu}_{k,s}(t) =  2 \cos^{2}(\theta) \sin^{2}(\theta)\bigg\{1 +
 \Re \bigg[ \frac{1}{\sqrt[2]{4m_1m_2t_0 t}} e^{-\frac{\pi k^2 (m_1+m_2)}{2m_1m_2a_0^2}}
  \bigg\{ W_{\kappa_2,\frac{1}{4}} (-2im_2t_0) W^{*}_{\kappa_1,\frac{1}{4}} (-2im_1t_0)  \\
  \nonumber && + \frac{4}{m_1 m_2 a_0^2 t_0}
 \left(W_{\kappa_2, \frac{1}{4}} (- 2 im_2t_0) - \frac{1}{8}\left(1 + \frac{ik^2}{m_2 a_0^2} \right)W_{\kappa_2 - 1,\frac{1}{4}} (-2 i m_2 t_0) \right)
 \left( W^{*}_{\kappa_1,\frac{1}{4}}(-2 im_1t_0) - \frac{1}{8} \left(1 + \frac{ik^2}{m_1 a_0^2} \right)W^{*}_{\kappa_1 -1,\frac{1}{4}}(-2 i m_1 t_0) \right) \bigg\}  \\
\nonumber && \times \bigg\{ W^{*}_{\kappa_2,\frac{1}{4}} (-2im_2t) W_{\kappa_1,\frac{1}{4}} (-2im_1t) \\
\nonumber && + \frac{4}{m_1 m_2 a_0^2 t}
 \left(W^{*}_{\kappa_2, \frac{1}{4}} (- 2 im_2t) - \frac{1}{8}\left(1 - \frac{ik^2}{m_2 a_0^2} \right)W^{*}_{\kappa_2 - 1,\frac{1}{4}} (-2 i m_2 t) \right)
 \left( W_{\kappa_1,\frac{1}{4}}(-2 im_1t) - \frac{1}{8} \left(1 - \frac{ik^2}{m_1 a_0^2} \right)W_{\kappa_1 -1,\frac{1}{4}}(-2 i m_1 t) \right) \bigg\}  \\
\nonumber && +  \frac{k^2}{\sqrt[2]{2m_1(2m_2)^3a_0^4t_0 t}} e^{-\frac{\pi k^2 (m_1+m_2)}{2m_1m_2a_0^2}}  \bigg\{ W_{\kappa_1,\frac{1}{4}} (-2im_1t_0) W^{*}_{-\kappa_2,\frac{1}{4}} (-2im_2t_0) \\
\nonumber && + \frac{1}{m_1 m_2 a_0^2 t_0}
 \left(W_{\kappa_1, \frac{1}{4}} (- 2 im_1t_0) - \frac{1}{8}\left(1 + \frac{ik^2}{m_1 a_0^2} \right)W_{\kappa_1 - 1,\frac{1}{4}} (-2 i m_1 t_0) \right)
 \left( W^{*}_{-\kappa_2,\frac{1}{4}}(2 im_2t_0) - \frac{2im_2a_0^2}{k^2} W^{*}_{-\kappa_2 +1,\frac{1}{4}}(2 i m_2 t_0) \right) \bigg\}  \\
\nonumber && \times \bigg\{ W^{*}_{\kappa_1,\frac{1}{4}} (-2im_1t) W_{-\kappa_2,\frac{1}{4}} (-2im_2t) \\
  && + \frac{1}{m_1 m_2 a_0^2 t}  \left(W^{*}_{\kappa_1, \frac{1}{4}} (- 2 im_1t) - \frac{1}{8}\left(1 - \frac{ik^2}{m_1 a_0^2} \right)W^{*}_{\kappa_1 - 1,\frac{1}{4}} (-2 i m_1 t) \right) \left( W_{-\kappa_2,\frac{1}{4}}(2 im_2t) + \frac{2im_2a_0^2}{k^2} W_{-\kappa_2 +1,\frac{1}{4}}(2 i m_2 t) \right) \bigg\} \bigg] \bigg\}
\end{eqnarray}
}

\subsection{Spacetimes with asymptotically flat regions}

 The FLRW spacetimes considered above are among the few non--trivial metrics for which the Dirac equation \eqref{DiracEquation} can be solved analytically. More often one does not have an exact solution at his disposal, and the implementation of eqs. \eqref{DefinitiveProbs} is a complicated task. In many cases of interest, however, the spacetime $M$ admits asymptotically flat regions $\Omega_A, \Omega_B \subset M$, usually in the far past and in the far future. $\Omega_A$ and $\Omega_B$ are separated by a region with non-trivial curvature, where the Dirac equation is usually unsolvable analytically. In $\Omega_A$ and $\Omega_B$ the solutions to the Dirac equation are the flat space modes $\{u^{I}_{k,s,i} (x) , v^{I}_{k,s,i}\}$  with $I=A,B$, and one has a natural choice for the positive frequency modes. Because of the non-trivial curvature, the two sets of solutions $A,B$ are distinct. When limited to one of the two regions, eqs. \eqref{DefinitiveProbs} reduce to the ordinary flat space formulae. When the intermediate curvature region is involved, a direct application of eqs. \eqref{DefinitiveProbs} is prohibitive. Nevertheless, if one is able to provide the relation between the two sets $A,B$, in the form of a Bogoliubov transformation, it is possible to derive oscillation formulae for the propagation from $\Omega_A$ to $\Omega_B$.
Assume that $\Omega_A = \bigcup_{\tau \leq \tau_A} \Sigma_{\tau}$ and  $\Omega_B = \bigcup_{\tau \geq \tau_B} \Sigma_{\tau}$ for some values $\tau_B > \tau_A$, and consider $\tau_0 \leq \tau_A$ , $\tau \geq \tau_B$. Suppose also that the $B$ modes are given in terms of the $A$ modes as
\begin{eqnarray}\label{BogoliubovModes2}
   u^{B}_{k',s',i} &=& \sum_{k,s} \left(\Gamma_{k' ,s'; k ,s; i}^{*} u^{A}_{k,s,i} + \Sigma_{k', s';k, s; i}^{*} v^{A}_{k, s,i}\right) \\
  v^{B}_{k' s',i} &=& \sum_{k,s}\left( \Gamma_{k', s'; k, s; i} v^A_{k,s,i} - \Sigma_{k', s';k ,s; i} u^A_{k,s,i} \right) \ .
\end{eqnarray}
Here the Bogoliubov coefficients $\Gamma_{k', s'; k, s; i}, \Sigma_{k', s';k ,s; i}$ are again provided by the inner products $(u^A_i,u^B_i),(u^A_i,v^B_i)$, yet their significance is slightly different from those in \eqref{BogoliubovModes}. While equation \eqref{BogoliubovModes} describes a general and arbitrary change of basis, the transformation of equation \eqref{BogoliubovModes2} is dictated by the circumstance of having a natural choice for the modes in $\Omega_A$ and $\Omega_B$, with a well--defined physical meaning. Indeed, the Bogoliubov coefficients take into account the effect of the curvature in the intermediate region. Then we can specialize equations \eqref{MixingCoefficient1}, \eqref{MixingCoefficient2} to get
\begin{eqnarray}\label{MixingCoefficient5}
 \nonumber && \Lambda^{B}_{q,r;k,s} (\tau) =   \sum_{q',k',r',s'} \bigg[ \Gamma_{q,r;q',r';2}  \left( \Gamma_{k,s;k',s';1}^{*} \Lambda_{q',r';k',s'}^{A}(\tau) -  \Sigma_{k,s;k',s';1}^{*} \Xi_{q',r';k',s'}^{A}(\tau) \right) \\
  &&+   \Sigma_{q,r;q',r';2} \left(\Gamma_{k,s;k',s';1}^{*} \Xi_{q',r';k',s'}^{A*} (\tau) +   \Sigma_{k,s;k',s';1}^{*} \Lambda_{q',r';k',s'}^{A*}(\tau) \right) \bigg] \ .
\end{eqnarray}
\begin{eqnarray}\label{MixingCoefficient6}
 \nonumber && \Xi^B_{q,r;k,s}(\tau) =  \sum_{q',k',r',s'} \bigg[\Gamma_{q,r;q',r';2}  \left( \Gamma_{k,s;k',s';1} \Xi^A_{q',r';k',s'}(\tau) +   \Sigma_{k,s;k',s';1} \Lambda^A_{q',r';k',s'}(\tau) \right) \\
  && -   \Sigma_{q,r;q',r';2}  \times \left(\Gamma_{k,s;k',s';1} \Lambda_{q',r';k',s'}^{A*} (\tau) - \Sigma_{k,s;k',s';1} \Xi_{q',r';k',s'}^{A*}(\tau) \right) \bigg] \ .
 \end{eqnarray}
 For $\tau < \tau_A$ $\Lambda^A(\tau),\Xi^A(\tau)$ are trivial, and vice-versa, for $\tau > \tau_B$, $\Lambda^B(\tau),\Xi^B(\tau)$ are trivial. Choosing, for instance, the $B$ representation, one would have a trivial expression for $\Lambda^B(\tau), \Xi^B(\tau)$ and one can make use of eqs. \eqref{MixingCoefficient5}, \eqref{MixingCoefficient6} to obtain $\Lambda^B(\tau_0), \Xi^B(\tau_0)$ in terms of $\Lambda^A(\tau_0),\Xi^A(\tau_0)$, which are also trivial. Then one can plug $\Lambda^B(\tau,\tau_0)$ and $\Xi^B (\tau, \tau_0)$ in equation \eqref{DefinitiveProbs} to obtain $P^{e \rightarrow \mu}_{k,s \rightarrow q,r}(\tau)$ for $\tau \geq \tau_B$ and the reference hypersurface $\tau_0 \leq \tau_A$.

\subsubsection{Scwharzschild black hole}

 As a realization of this scheme, consider the (static) Schwarzschild metric
\begin{eqnarray}
 ds^2 = \left(1 - \frac{2GM}{r}\right)dt^2 -\left(1 - \frac{2GM}{r}\right)^{-1}dr^2 - r^2 d\Omega ,
  \end{eqnarray}
  and two sequences of spacelike hypersurfaces \cite{Note6} $\{\Sigma_{n}^{+}\}_{n \in \mathbb{N}}$, $\{\Sigma_{n}^{-}\}_{n \in \mathbb{N}}$ approaching, respectively, the future $\mathcal{I}^{+}$ and the past null infinity $ \mathcal{I}^{-}$ as $n \rightarrow \infty$. We require that $\Sigma_{n}^{+}$ and $\Sigma_{n}^{-}$ be Cauchy surfaces respectively for the causal past $J^{-}(\mathcal{I}^{+})$ of $\mathcal{I}^{+}$  and the causal future $J^{+} (\mathcal{I}^{-})$of $\mathcal{I}^{-}$ for each $n$. For $n$ large enough, these surfaces span an approximately flat portion of the Schwarzschild spacetime.  On the surfaces $\Sigma_{n}^{-}$, as $n$ approaches infinity, we expand the massive fields in terms of the incoming solutions, with frequency defined with respect to the schwarzschild time $t$ and with definite angular momentum,

  \begin{eqnarray}
    \zeta^{IN}_{\omega,\kappa_j,m_j;i} (t,r,\theta,\phi) \propto e^{-i \omega t}\,, \qquad \qquad \xi^{IN}_{\omega,\kappa_j,m_j;i} (t,r,\theta, \phi) \propto e^{i \omega t}.
  \end{eqnarray}

      These modes reduce to the flat space solutions \eqref{inmodes1} and \eqref{inmodes2} as $\mathcal{I}^{-}$ is approached. Omitting the irrelevant angular and spin quantum numbers, as $n \rightarrow \infty$, we get
  \begin{eqnarray}
  \Lambda^{IN}_{\omega;\omega'} (\Sigma^{-}_n) \rightarrow \delta_{\omega', \sqrt{\omega^2 + \Delta m^2} }|U_{\omega,\omega'}|e^{i \varphi^{-}(\omega,n)}
  \\
  \Xi^{IN}_{\omega;\omega'} (\Sigma^{-}_n) \rightarrow \delta_{\omega', \sqrt{\omega^2 + \Delta m^2} }|U_{\omega,\omega'}|e^{i \rho^{-}(\omega,n)}
  \end{eqnarray}
  with $|U_{\omega,\omega'}|$ and $|V_{\omega,\omega'}|$  flat space spherical mixing coefficients as defined in \eqref{MixingCoeffSpher1} and \eqref{MixingCoeffSpher2}, and $\varphi^{-} (\omega,n), \rho^{-} (\omega,n)$ phase factors depending on $\omega$ and $n$. A similar reasoning can be carried on for the \emph{outgoing} modes emerging at $\mathcal{I}^+$, $\zeta^{OUT}_{\omega,\kappa_j,m_j;i} (t,r,\theta,\phi) $ , $\xi^{OUT}_{\omega,\kappa_j,m_j;i} (t,r,\theta, \phi)$, so to yield as $n \rightarrow \infty$
      \begin{eqnarray}
   \Lambda^{OUT}_{\omega;\omega'} (\Sigma^{+}_n) \rightarrow \delta_{\omega', \sqrt{\omega^2 + \Delta m^2} } |U_{\omega,\omega'}|e^{i \varphi^{+}(\omega,n)}
   \\
    \Xi^{OUT}_{\omega;\omega'} (\Sigma^{+}_n) \rightarrow \delta_{\omega', \sqrt{\omega^2 + \Delta m^2} } |V_{\omega,\omega'}|e^{i \rho^{+}(\omega,n)}\,.
     \end{eqnarray}
Because of the Black Hole, the $IN$ and $OUT$ modes do not coincide. Fermion creation by the Schwarzschild black hole has been studied, via the tunneling method, in \cite{Kerner}. There it has been shown that the Hawking temperature $T_H = \frac{1}{8\pi G M}$ is recovered for the emission of spin--$ \frac{1}{2}$ particles. We then infer that the $IN$ and $OUT$ modes are related by a thermal Bogoliubov transformation at the Hawking temperature $T_H$, corrected for fermions:

\begin{eqnarray}\label{BogoliubovHawking}
  && \zeta^{OUT}_{\omega,\kappa_j,m_j;i} =  \sqrt{\frac{e^{\frac{\omega}{k_B T_H}}}{e^{\frac{\omega}{k_B T_H}}+1}} \zeta^{IN}_{\omega,\kappa_j,m_j;i} + \sqrt{\frac{1}{e^{\frac{\omega}{k_B T_H}}+1}} \xi^{IN}_{\omega,\kappa_j,m_j;i}
 \end{eqnarray}
 \vspace{0.1cm}
 \begin{eqnarray}
   && \xi^{OUT}_{\omega,\kappa_j,m_j; i} = \sqrt{\frac{e^{\frac{\omega}{k_B T_H}}}{e^{\frac{\omega}{k_B T_H}}+1}}\xi^{IN}_{\omega,\kappa_j,m_j;i}- \sqrt{\frac{1}{e^{\frac{\omega}{k_B T_H}}+1}} \zeta^{IN}_{\omega,\kappa_j,m_j;i}  \ ,
\end{eqnarray}
It is understood that these equations hold as long as the spacetime is stationary (eternal black hole). For a body that collapses to a Schwarzschild black hole, as considered in the original paper by Hawking \cite{Hawking}, we expect a slightly modified version of the Bogoliubov transformations, comprising non--diagonal thermal coefficients
$\Gamma_{\omega, \omega';i}$ and $\Sigma_{\omega, \omega';i}$ with $\omega \neq \omega'$.
We pick the ingoing representation to calculate the probabilities (of course, the outgoing representation yields the same result, as the Bogoliubov transformations are diagonal), and employ eqs. (\ref{MixingCoefficient5},~\ref{MixingCoefficient6}) to obtain, with $F_H (\omega) = \frac{1}{e^{\frac{\omega}{k_B T_H}}+1}$,

 \begin{eqnarray}
  \nonumber \Lambda^{IN}_{\omega;\omega'} (\Sigma^{+}_n) &=& \bigg\{ \sqrt{\left[1 - F_H (\omega)\right] \left[1 - F_H (\omega')\right]}  \Lambda^{OUT}_{\omega;\omega'} (\Sigma^{+}_n)   - \ \ \sqrt{F_H(\omega)\left[1 - F_H(\omega') \right]} \Xi^{OUT}_{\omega;\omega'} (\Sigma^{+}_n)   \\   &+& \sqrt{F_H(\omega')\left[1 - F_H(\omega) \right]} \left(\Xi^{OUT}_{\omega;\omega'}\right)^{*} (\Sigma^{+}_n)   + \sqrt{F_H(\omega)F_H(\omega')} \left(\Lambda^{OUT}_{\omega;\omega'}\right)^{*} (\Sigma^{+}_n)  \bigg\}  \,,
 \end{eqnarray}
 Considered that \mbox{$\Lambda^{OUT}_{\omega;\omega'} (\Sigma^{+}_n) \rightarrow \delta_{\omega', \sqrt{\omega^2 + \Delta m^2} }  |U_{\omega,\omega'}|e^{i \varphi^{+}(\omega,n)}$} and \mbox{$\Xi^{+}_{\omega;\omega'} (\Sigma^{+}_n) \rightarrow \delta_{\omega', \sqrt{\omega^2 + \Delta m^2} } |V_{\omega,\omega'}|e^{i \rho^{+}(\omega,n)}$} as $n \rightarrow \infty$, we obtain

 \begin{eqnarray}\label{Hawking1}
 \nonumber  \Lambda^{IN}_{\omega;\omega'} (\Sigma^{+}_n) &=& \bigg\{ \sqrt{\left[1 - F_H (\omega)\right] \left[1 - F_H (\omega')\right]} |U_{\omega;\omega'}|e^{i \varphi^{+}(\omega,n)}  - \sqrt{F_H(\omega)\left[ 1 - F_H(\omega') \right]} |V_{\omega;\omega'}| e^{i \rho^{+}(\omega,n)}  \\
    &+& \sqrt{F_H(\omega')\left[ 1 - F_H(\omega) \right]} |V_{\omega;\omega'}| e^{-i \rho^{+}(\omega,n)} + \sqrt{F_H(\omega)F_H(\omega')} |U_{\omega,\omega'}|e^{-i \varphi^{+}(\omega,n)} \bigg\}\delta_{\omega', \sqrt{\omega^2 + \Delta m^2} }
 \end{eqnarray}
 and

 \begin{eqnarray}\label{Hawking2}
  \nonumber \Xi^{IN}_{\omega;\omega'} (\Sigma^{+}_n) &=& \bigg\{\sqrt{\left[1 - F_H (\omega)\right] \left[1 - F_H (\omega')\right]} |V_{\omega,\omega'}|e^{i \rho^{+}(\omega,n)}  + \sqrt{F_H(\omega)\left[1 - F_H(\omega') \right]} |U_{\omega,\omega'}| e^{i \varphi^{+}(\omega,n)}  \\
  && - \sqrt{F_H(\omega')\left[1 - F_H(\omega) \right]} |U_{\omega,\omega'}| e^{- i \varphi^{+}(\omega,n)}   + \sqrt{F_H(\omega)F_H(\omega')} |V_{\omega,\omega'}|e^{-i \rho^{+}(\omega,n)}\bigg\} \delta_{\omega', \sqrt{\omega^2 + \Delta m^2} } \ .
 \end{eqnarray}
 Choosing as reference hypersurface $\Sigma^{-}_m$ for large $m$, we can now compute the probabilities \eqref{DefinitiveProbs} for a neutrino propagating from $\Sigma^{-}_m$ to $\Sigma^{+}_{n}$, i.e. from $\mathcal{I}^{-}$ to $\mathcal{I}^{+}$ in the limit $m,n \rightarrow \infty$. We find, for $m,n \rightarrow \infty$
\begin{eqnarray}\label{HawkingProb}
\nonumber    P^{e \rightarrow \mu}_{\omega} (m,n) &\approx&  2 \cos^2\theta \sin^2 \theta \
\bigg( 1   - \sqrt{\left[1 - F_H (\omega)\right] \left[1 - F_H (\omega')\right]}   \left[|U_{\omega;\omega'}|^2 \cos (\Delta^{-}_{\omega;m,n}) + |V_{\omega;\omega'}|^2 \cos(\Phi^{-}_{\omega;m,n})\right]  \\
\nonumber  &+& \sqrt{F_H(\omega)\left[1 - F_H(\omega') \right]} |U_{\omega;\omega'}| |V_{\omega;\omega'}|   \left[ \cos (\Theta^{-}_{\omega;m,n}) - \cos(\Psi^{-}_{\omega;m,n})\right] \\
\nonumber &+& \sqrt{F_H(\omega')\left[1 - F_H(\omega) \right]}|U_{\omega;\omega'}| |V_{\omega;\omega'}|  \left[ \cos(\Psi^{+}_{\omega;m,n})- \cos(\Theta^{+}_{\omega;m,n})\right]  \\
  &-& \sqrt{F_H(\omega)F_H(\omega')}  \left[|U_{\omega;\omega'}|^2\cos(\Delta^{+}_{\omega;m,n}) + |V_{\omega;\omega'}|^2 \cos(\Phi^{+}_{\omega;m,n})  \right] \bigg)  \ .
\end{eqnarray}
with $\Delta^{\pm}_{\omega;m,n}= \varphi^{+} (\omega,n) \pm \varphi^{-} (\omega,m)$ , $\Phi^{\pm}_{\omega;m,n} = \rho^{+} (\omega,n) \pm \rho^{-} (\omega,m)$, $\Psi^{\pm}_{\omega;m,n} = \rho^{+} (\omega,n) \pm \varphi^{-} (\omega,m)$  and $\Theta^{\pm}_{\omega;m,n} = \varphi^{+} (\omega,n) \pm \rho^{-} (\omega,m)$ . In particular, for large energies of the mass fields $\omega,\omega'$, $|V_{\omega,\omega'}|\rightarrow 0$ and $|U_{\omega,\omega'}| \rightarrow 1$, thus
\begin{eqnarray}\label{HawkingProbLargeMomentum}
     P^{e \rightarrow \mu}_{\omega} (m,n) \approx  2 \cos^2\theta \sin^2 \theta  \  \bigg( 1 - \sqrt{\left[1 - F_H (\omega)\right] \left[1 - F_H (\omega')\right]} \cos (\Delta^{-}_{\omega,\kappa_j;m,n})
- \sqrt{F_H(\omega)F_H(\omega')} \cos(\Delta^{+}_{\omega,\kappa_j;m,n})\bigg) \ ,
\end{eqnarray}
where it is understood that $\omega^{' 2} = \omega^2 + \Delta m^2$. If the limit $T_H \rightarrow 0$ is taken in equation \eqref{HawkingProbLargeMomentum}, one obtains, apart from a phase factor, the usual Pontecorvo oscillation formulae.
To compute the oscillation formulae on a Schwarzschild background for an arbitrary propagation, since exact analytical solutions are unavailable, one has to resort to approximate solutions to the Dirac equation.
In all the applications considered, the oscillation formulae do not depend on the helicity $s$ of neutrinos. However, when additional complications due to frame--dragging and non--conservation of angular momentum arise (see e.g. \cite{Lambiase1,Lambiase2}), the formulae for left-handed and right-handed neutrinos can differ.

\subsection{Quantum mechanical limit}

It is a general feature of equations \eqref{DefinitiveProbs} that when all the quantum field theoretical effects are negligible, the oscillation formulae are modified only for a phase factor. Indeed, when one can neglect $\Xi_{k,s}$ in equation \eqref{DefinitiveProbs}, one immediately has $|\Lambda^{*}_{k,s}(\tau) | =1$, ($|\Lambda_{\omega,\omega'} (\tau)| = 1$ respectively, in the non--diagonal case) for each $\tau$ from equation \eqref{Compatibility}. Then the product $\Lambda_{k,s}(\tau_0)^* \Lambda_{k,s} (\tau)$, ($\Lambda_{\omega,\omega'}(\tau_0)^* \Lambda_{\omega,\omega'} (\tau)$ respectively) is just a phase $e^{i \varphi (\tau_0,\tau)}$ and the net effect is a phase shift with respect to the Pontecorvo oscillation formulae, consistently with previous results obtained in a heuristic treatment\cite{Cardall}. Of course, the explicit value of the phase \mbox{$\varphi (\tau_0,\tau)$} depends on the metric and the surfaces $\tau$ considered, as well as on the mode expansion chosen for the mass fields. When the gravitational fields are weak enough, the phase can be computed by means of geometrical optics considerations \cite{Cardall,Lambiase1}.

\section{Conclusions}

We have developed a quantum field theoretical approach to the vacuum neutrino oscillations in curved space, discussing the transition probabilities, and their behaviour under changes of mass representation. We have analyzed the non--trivial interplay between quantum field mixing and field quantization in curved space, and have found that the former has a remarkably richer structure when compared to its flat space counterpart. In particular, the formalism has to be versatile enough to deal with the existence of infinite unitarily inequivalent representations of the mass fields, which is to say that no preferred notion of particle does generally exist. In the spirit of general covariance, we have determined the effect on flavor fields of a shift in the expansion of the mass fields, and established under which conditions the resulting transition probabilities are the same. We have then computed the oscillation formulae in three example metrics, including two FRLW spacetimes and the static Schwarzschild black hole. In the latter we have found that the Hawking radiation affects, although very slightly, the oscillations for neutrinos propagating from the asymptotic past infinity to the asymptotic future infinity. As a general result, it is found that when all the quantum field theoretical effects on neutrino mixing can be neglected, the gravitational background only affects the phase of the oscillations, constistently with previous analyses.

\section*{Acknowledgements}
Partial financial support from MIUR and INFN is acknowledged.
A.C. and G.L. also  acknowledge the COST Action CA1511 Cosmology
and Astrophysics Network for Theoretical Advances and Training Actions (CANTATA).


\begin{thebibliography}{99}

\bibitem{Pauli}
L. M. Brown, \textit{Physics Today} {\bf 31}, 9, 23 (1978).

\bibitem{NeutrinoOscillations1}
 Q.R. Ahmad et al. (SNO Collaboration), \textit{Phys. Rev. Lett.} {\bf 87} (7): 071301 (2001).

\bibitem{NeutrinoOscillations2}
Y. Fukuda et al. (Super-Kamiokande Collaboration), \textit{Phys. Rev. Lett.} {\bf 81} (8), 1562 - 1567 (1998).

\bibitem{Pontecorvo}
S. M. Bilenky and B. Pontecorvo, \textit{Phys. Rep.} {\bf 41}, 225 (1978).

\bibitem{Bilenky}
S. M. Bilenky and S. T. Petcov, \textit{Rev. Mod. Phys.} {\bf 59}, 671 (1987)

\bibitem{Mohapatra}
R. N. Mohapatra and P. B. Pal,  \textit{Massive neutrinos in physics and astrophysics},  \textit{Lecture Notes in Physics} (3rd ed.) {\bf{72}}, (2007).

\bibitem{NeutrinoMass}
Y. Cai et al. \textit{From the Trees to the Forest: A Review of Radiative Neutrino Mass Models}, \textit{Front. Phys.} {\bf{5}},
10.3389/fphy.2017.00063  (2017).

\bibitem{NeutrinoNature}
S. T. Petcov, \textit{Adv. High En. Phys.}, 852987 (2013)

\bibitem{Franckowiak}
A. Franckowiak, \textit{J. Phys.: Conf. Ser.} {\bf{888}}, 012009 (2017).

\bibitem{Buchmuller}
W. Buchm{\"u}ller, \textit{Neutrinos, Grand Unification and Leptogenesis}, arXiv:hep-ph/0204288v2 (2002).

\bibitem{CosmologicalNeutrinos}
M. Tanabashi et al. (Particle Data Group), \textit{Neutrinos in Cosmology}, \textit{Phys. Rev. D} {\bf 98}, 030001 (2018).

\bibitem{PTOLEMY}
M.G. Betti et al. (PTOLEMY collaboration), \textit{JCAP} 07, 047 (2019).

\bibitem{Mavans}
D. B. Kaplan, A. E. Nelson, N. Weiner, \textit{Phys. Rev. Lett.} {\bf 93}, 091801 (2004);
R. Fardon, A. E. Nelson, N. Weiner, \textit{JCAP}  \textbf{10}, 005 (2004).

\bibitem{Grossman}
Y. Grossman and H. J. Lipkin, \textit{Phys. Rev. D} \textbf{55}, 2760 (1997).

\bibitem{Piriz}
D. Piriz, M. Roy and J. Wudka, \textit{Phys. Rev. D} \textbf{54}, 1587 (1996).

\bibitem{Cardall}
C. Y. Cardall and G. M. Fuller, \textit{Phys. Rev. D} \textbf{55}, 7960 (1997).

\bibitem{Birrell}
N. Birrell and P. Davies,  \textit{Quantum Fields in Curved Space (Cambridge Monographs on Mathematical Physics).}, Cambridge: Cambridge University Press,  doi:10.1017/CBO9780511622632 (1982).

\bibitem{Wald}
R. Wald, \textit{Quantum Field Theory in Curved Spacetime and Black Hole Thermodynamics (Chicago Lectures in Physics)}, The University of Chicago Press, ISBN: 9780226870274 (1994).

\bibitem{Note1}
Here the index $k$ refers to a generic set of quantum numbers labelling the modes. The sums $\sum_{k}$ are understood as integrals if $k$ is a continuous index.

\bibitem{Capolupo1}
M.~Blasone, A.~Capolupo, G.~Vitiello, {\it  Phys.\ Rev.\ D} {\bf 66}, 025033 (2002) and references therein;
A.~Capolupo, I.~De Martino, G.~Lambiase and A.~Stabile,
	Phys.\ Lett.\ B {\bf 790}, 427 (2019);
	%
	%
K. Fujii, C. Habe and T. Yabuki, Phys. Rev. D {\bf 59}, 113003 (1999); Phys. Rev. D {\bf 64}, 013011 (2001);
K.C. Hannabuss and D.C. Latimer, J. Phys. A {\bf 33}, 1369 (2000); J. Phys. A 36, L69 (2003);
C.R Ji and Y. Mishchenko, Phys. Rev. D {\bf 65}, 096015 (2002); Ann. Phys. {\bf 315}, 488 (2005);
	M. Blasone, A. Capolupo, O. Romei and  G. Vitiello,
	Phys.\ Rev.\ D  {\bf 63}, 125015  (2001);
	%
	A. Capolupo, C. R. Ji ,  Y. Mishchenko and  G. Vitiello,
	Phys.\ Lett.\ B  {\bf 594}, 135 (2004);
	%
	M.~Blasone, A.~Capolupo, F.~Terranova, G.~Vitiello,
	Phys.\ Rev.\ D {\bf 72}, 013003 (2005).


\bibitem{Capolupo:2006et}
	A.~Capolupo,
	Adv. High Energy Phys.  {\bf 2016},   8089142, 10 (2016);
	Adv.\ High Energy Phys.\  {\bf 2018}, 9840351 (2018);
	%
	A. Capolupo, S. Capozziello and G. Vitiello,
	Phys.\ Lett.\ A  {\bf 373}, 601 (2009);
	%
	%
	Phys.\ Lett.\ A  {\bf 363}, 53 (2007);
	%
	Int.\ J.\ Mod.\ Phys.\ A  {\bf 23}, 4979 (2008);
	%
	M.~Blasone, A.~Capolupo, S.~Capozziello, G.~Vitiello,
	Nucl.\ Instrum.\ Meth.\  A  {\bf588}, 272 (2008);
	%
	M.~Blasone, A.~Capolupo, G.~Vitiello,
	Prog.\ Part.\ Nucl.\ Phys.\   {\bf 64}, 451 (2010);
	M. Blasone,  A. Capolupo,  S. Capozziello,  S. Carloni and  G. Vitiello,
	Phys.\ Lett.\ A   {\bf 323}, 182 (2004);
	%
	A.~Capolupo, M.~Di Mauro, A.~Iorio,
	Phys.\ Lett.\ A  {\bf 375}, 3415 (2011).


\bibitem{Capolupo:2019gmn}
  A.~Capolupo, S.~M.~Giampaolo, G.~Lambiase and A.~Quaranta,
  arXiv:1912.03332 [hep-ph].



\bibitem{Note3}
The helicity index can always be decoupled by choosing eigenstates of the helicity operator.

\bibitem{Note4}
This does \emph{not} imply that $Q_e$ and $Q_{\mu}$ are conserved separately.

\bibitem{AngularMom}
 L. Biedenharn, J. Louck, P. Carruthers, \textit{Angular Momentum in Quantum Physics: Theory and Application (Encyclopedia of Mathematics and its Applications)}, Cambridge: Cambridge University Press., doi:10.1017/CBO9780511759888  (1984).

\bibitem{Thaller}
B. Thaller, \textit{Advanced Visual Quantum Mechanics}, Springer--Verlag New York, doi:10.1007/b138654 (2005).

\bibitem{Abramowitz}
M. Abramowitz and I. A. Stegun, \textit{Handbook of Mathematical Functions}, Dover Publications Inc. New York, ISBN: 9780486612720 (1965).



\bibitem{Note5}
In this case the label $\omega$ refers to the energies of the reference mass field, that is $\psi_1$ for $\rho = e$ and $\psi_2$ for $\rho = \mu$. This \emph{must not} be interpreted as the energy of neutrinos with flavor $\rho$, since by definition they have no definite energy. Indeed, the non--diagonality of \eqref{MixingCoeffSpher} shows that it is impossible to classify flavor neutrinos with a single energy $\omega$.

\bibitem{Barut}
A. O. Barut, H. Duru, \textit{Phys. Rev. D}  {\bf 36}, 3705 (1987).

\bibitem{Note6}
We use here two discrete families of surfaces, but also continuous families would do the trick, as long as the same limiting process is involved ($\Sigma^{\pm} \rightarrow \mathcal{I}^{\pm}$).

\bibitem{Kerner}
R. Kerner, R. B. Mann, \textit{Class.Quant.Grav.} \textbf{25}, 095014 (2008).

\bibitem{Hawking}
S. W. Hawking, \textit{Particle creation by black holes}, \textit{Comm. Math. Phys.}  \textbf{43} , no. 3, 199--220 (1975).

\bibitem{Lambiase1}
G. Lambiase, G. Papini, Raffaele Punzi, G. Scarpetta, \textit{Phys.Rev. D} \textbf{71}, 073011 (2005).

\bibitem{Lambiase2}
G. Lambiase, \textit{Mon.Not.Roy.Astron.Soc.} 362, 867-871 (2005).


\end{thebibliography}
\end{document}